\documentstyle[floats,preprint,eqsecnum,prd,aps,epsf]{revtex}
\begin{document}

\def\grtsim{\,\,\rlap{\raise 3pt\hbox{$>$}}{\lower 3pt\hbox{$\sim$}}\,\,}
\def\lsim{\,\,\rlap{\raise 3pt\hbox{$<$}}{\lower 3pt\hbox{$\sim$}}\,\,}

\input FEYNMAN

\def\grtsim{\,\,\rlap{\raise 3pt\hbox{$>$}}{\lower 3pt\hbox{$\sim$}}\,\,}
\def\lsim{\,\,\rlap{\raise 3pt\hbox{$<$}}{\lower 3pt\hbox{$\sim$}}\,\,}

\draft
 
{\tighten
\preprint{\vbox{\hbox{CALT-68-2156}
                \hbox{hep-ph/9802335} 
		\hbox{\footnotesize DOE RESEARCH AND}
		\hbox{\footnotesize DEVELOPMENT REPORT}
                 }}
 
\title{Diffusion and Decoherence of Squarks and Quarks During the
Electroweak Phase Transition}
 
\author{Hooman Davoudiasl\footnote{E-mail address:
hooman@theory.caltech.edu} and Eric Westphal\footnote{E-mail address:
westphal@theory.caltech.edu}}
 
\address{California Institute of Technology, Pasadena, CA 91125 USA}

\maketitle 

\begin{abstract} 

To estimate the diffusion constant $D$ of particles in a plasma, we develop a
method that is based on the mean free path $\lambda$ for scatterings with
momentum transfer $q \grtsim T$.  Using this method, we estimate $\lambda$ and
$D$ for squarks and quarks during the electroweak phase transition.  Assuming
that Debye and magnetic screening lengths provide suitable infrared cutoffs,
our calculations yield $\lambda \lsim 18/T$ and $D \lsim 5/T$ for both squarks
and quarks.  Our estimate of $\lambda$ suggests that suppressions of charge
transport due to decoherence of these strongly interacting particles during
the electroweak phase transition are not severe and that these particles may
contribute significantly to electroweak baryogenesis.

\end{abstract}  

}

\newpage

\pagenumbering{arabic} 

\section{Introduction}

In recent years, many scenarios incorporating extensions of the
Minimal Standard Model (MSM) have been proposed for the generation of
the Baryon Asymmetry of the Universe during the electroweak phase
transition.  A much studied mechanism uses the transport of some
quantum number, via the CP violating interactions of certain particles
with the expanding broken phase Higgs wall, into the unbroken phase
during a first order electroweak phase transition.  This transported
quantum number, such as the axial top quark number, then biases the
equilibrium in the direction of baryon number violation, ultimately
generating a baryon asymmetry through weak sphaleron
processes\cite{CKN,HN}.  Of the extensions of the MSM used in these
scenarios, those based on supersymmetric models such as the Minimal
Supersymmetric Standard Model (MSSM) are the best motivated.  By
including supersymmetric particles, extra CP violation can be
naturally introduced into a mechanism for baryogenesis.  It is argued
that a strongly first order phase transition can be achieved if the
broken phase right-handed stop mass is less than, or of the order of,
the top quark mass \cite{Carena,Riotto}.  In this case, squarks with
soft supersymmetry breaking masses $m_s \sim T$, where $T \sim 100$
GeV is the temperature of the plasma, could have an important role in
the charge transport mechanism\cite{DRW}.  However, to study
charge transport via stops, one needs to have an estimate of their
diffusion properties in a plasma at the electroweak scale.

In Refs.\ \cite{JPT,Moore}, a set of approximations in conjunction with 
the Boltzmann equation for quarks in the
plasma of MSM particles were used to estimate the quark diffusion
constant $D_q$.  As strong interactions dominate the diffusion
process, and since stops are strongly interacting particles, it has
been assumed that the estimate $D_q \sim 6/T$ of Ref.\
\cite{JPT} is applicable to stops as well (the validity of this
assumption is not {\it a priori} obvious because of the different
statistics, masses, and couplings of squarks and quarks).  This
estimate is derived using an approximate gluon propagator with the
thermal mass of the longitudinal gluons $m_g$ as an infrared cutoff
and ignoring the different thermal properties of the transverse and
longitudinal gluons.  The use of the approximate gluon propagator can
only yield the leading-logarithmic behavior and does not result in the
correct leading $\alpha_s^2$ non-logarithmic contribution\cite{JPT,Moore}.
However, the diffusive process is expected to be dominated by the
$t$-channel gluon exchange diagrams, and for these diagrams, the
leading logarithm contribution is expected to be dominant\cite{JPT}.
A more comprehensive treatment in Ref.\ \cite{Moore} gives $D_q \sim
3/T$.  Again, this estimate is at the level of the leading logarithm.

To study the diffusion of particles in a plasma, one must consider scattering
processes in which the momentum transfer $q$ is not small.  For the
$t$-channel processes we consider in this paper, we approximate this effect by
an infrared regularization of $q$ such that $q \grtsim T$, since the typical
momenta of the scattering particles are of order $T$.  That is, we assume that
such a transfer of momentum in the scattering process randomizes the momentum
of the diffusing particle, as an approximation of the physics invovled.  This
momentum randomization approximation is implemented naturally by using as
cutoffs the longitudinal and transverse thermal masses of the exchanged gluons
which are comparable in magnitude to the temperature $T$ of the plasma.  We
note here that the thermal masses of gluons depend parametrically on the
strong coupling constant $g_s$, and that these masses are comparable to the
temperature only for realistic values of $g_s \sim 1$.  If we take the limit
in which $g_s \ll 1$, these masses will be small compared to the temeprature
$T$, and cannot be used as cutoffs in our approach.  In this paper, we use the
above momentum randomization approximation to calculate the
elastic\footnote{Non-elastic processes in which the species of the particle
changes do contribute to the mean free path; however, these are not considered
to be diffusive processes here.} mean free path $\lambda$ associated with the
diffusive processes (for which $q \grtsim T$)  and relate it to the diffusion
constant $D$.  Henceforth, the words ``mean free path'' refer to this
diffusive mean free path.  

In electroweak baryogenesis scenarios that use charge transport, the CP
violating interactions of the charge carriers with an expanding Higgs wall
eventually result in the generation of baryon number.  However, within the
Higgs wall, multiple scatterings in which the final momentum of the charge
carrier differs significantly from its initial momentum wash out the asymmetry
caused by CP violation and suppress baryogenesis.  This effect is known as
decoherence.  These same processes also contribute to the diffusion of the
particles within the plasma.  Therefore, the diffusive mean free path $\lambda$
which we calculate is a relevant parameter for estimating the supression due
to decoherence.

In this paper, we use the method described above to estimate the mean free path
$\lambda_s$ and the diffusion constant $D_s$ of stops that have a soft
supersymmetry breaking mass $m_s \sim T$ in the unbroken phase of the
electroweak plasma.  We use the same method to estimate the mean free path
$\lambda_q$ and the diffusion constant $D_q$ for quarks and compare our values
with those of Ref.\ \cite{JPT}.  We find that our method reproduces the results
of Ref.\ \cite{JPT} for the set of parameters used therein.  In general, our
results suggest that the values of $\lambda$ and $D$ of squarks are close to
those of quarks.

In calculating $\lambda$ and $D$, we consider only strong interactions, for
they dominate the diffusion of squarks and quarks in the plasma.  In the case
of squarks, we further assume that scatterings from and via the heavy gluinos
and squarks do not contribute significantly, leaving only quarks and gluons as
the dominant scatterers and mediators.  As consideration of more scatterers
can only decrease the calculated values of $\lambda$ and therefore $D$, the
inclusion of only quark scatterers will yield an upper bound (and even a
reasonable order of magnitude estimate) for the size of the squark diffusion
constant $D_s$, in light of the results of Ref.\ \cite{Moore}.  In computing
$D_q$, to facilitate comparison, we follow Ref.\ \cite{JPT} and only consider
$t$-channel quark-quark scattering.

As explained above, we implement our momentum randomization approximation using
the longitudinal and transverse gluons thermal masses $m_g$ and $m_t$ in the
plasma, referred to as Debye and magnetic masses, respectively, as physical
infrared cutoffs for the exchanged gluon momentum.  Since these masses depend
on $g_s$, the momentum transfer $q$ in our calculations is parametrically of
order $g_s T$, and the use of Hard Thermal Loop (HTL) propagator for the gluons
is valid\cite{Weldon,Braaten}.  In the limit $g_s \ll 1$, this corresponds to
small momentum transfer.  However, we note that in our approximation, we demand
$ q \grtsim T$.  Thus, in the $g_s \ll 1$ limit, in our approach, we have to
abandon the thermal masses as cutoffs, and choose a cutoff $q_{cut}$
parametrically independent of $g_s$, and such that $q_{cut} \grtsim T$.  Then,
in this technical sense, in our approach, the use of the HTL gluon propagator
would not be justified, since the momentum carried by the exchanged gluon would
not be parametrically soft, that is ${\cal O}(g_s T)$, as required by the HTL
formalism\cite{Weldon,Braaten}.  Our method does not yield the leading logarithm
behavior, as $g_s \to 0$, obtained in Refs. \cite{JPT,Moore} for the
$t$-channel processes considered therein.  Nonetheless, for the physical
values of $g_s$ at the electroweak phase transition, we obtain similar results,
suggesting that our scheme reasonably approximates the physics.  For realistic
values of $g_s$, we use the gluon thermal masses as cutoffs in our momentum
randomization approximation, since they naturally arise in the plasma.  Then,
parametrically, we are allowed to use the HTL propagator for the gluon.  In
this work, to incorporate thermal effects qualitatively, we approximate the
effect of the HTL propagator by separating the gluon propagator into transverse
and longitudinal parts that in general have different thermal masses.  Whereas
$m_g$ is calculable at one loop, $m_t$ is not calculable perturbatively and is
unknown.  Therefore, we will present our results for two representative values
of $m_t$.  

In the next section, we describe our method for calculating the diffusion
constant of particles in the plasma.  In Section III, we present our estimates
for $\lambda_s$, $D_s$, $\lambda_q$, and $D_q$, followed by a discussion of
our results.  The appendix contains some information on the approximate
thermal gluon propagator we use in our calculations.

\section{Calculation of the Mean Free Path and the Diffusion Constant}

Let us consider a two body scattering process where the initial and
final particles have 4-momenta $(p, k)$ and $(p^\prime, k^\prime)$,
respectively.  We refer to each particle by its 4-momentum for the
rest of this section.  The $p$-particle, whose diffusion constant we
calculate, scatters from the $k$-particle.  For processes relevant to
the calculation of the diffusion constant $D$, the final state
$p^\prime$-particle is of the same species as the initial
$p$-particle.  The
$p$-particle, $k$-particle, $p^\prime$-particle, and
$k^\prime$-particle have thermal distributions $\rho_p$, $\rho_k$,
$\rho_{p^\prime}$, and $\rho_{k^\prime}$, respectively.  The density
per unit volume of a particle with 4-momentum $p$ is given by $\rho_p
\, d^3 p/(2\pi)^3$.

The transition probability for the above process per unit volume and
per unit time is
\[
\eta = {d^3 p \over (2\pi)^3 \, 2 p^0}{d^3 k \over (2\pi)^3 \, 2 k^0}
{d^3 p^\prime \over (2\pi)^3 \, 2 p^{\prime 0}}
{d^3 k^\prime \over (2\pi)^3 \, 2 k^{\prime 0}}
(2 \pi)^4 \delta^{(4)}(p + k - 
p^\prime - k^\prime) |{\cal M}|^2 \] 
\begin{equation}
\times \rho_p \, \rho_k\,  (1 \pm \rho_{p^\prime}) \, 
(1 \pm \rho_{k^\prime}), 
\label{eta}
\end{equation}
where $\cal M$ is the amplitude for the scattering,
and $\pm$ is for final state bosons or fermions, respectively.
Let $d\sigma$ be the differential cross section for this process,
\begin{equation}
d\sigma = {d^3 p^\prime \over (2\pi)^3 \, 2 p^{\prime 0}} 
{d^3 k^\prime \over (2\pi)^3 \, 2 k^{\prime 0}} 
(2 \pi)^4 \delta^{(4)}(p + k - p^\prime - k^\prime) {|{\cal M}|^2 
(1 \pm \rho_{p^{\prime}}) (1 \pm \rho_{k^{\prime}})
\over 4 \sqrt {(p\cdot k)^2 - m_p^2 \, m_k^2}}.
\label{dsig}
\end{equation}
Comparing Eqs.\ (\ref{eta}) and (\ref{dsig}) yields
\begin{equation}
\eta = \left[{d^3 p \over (2\pi)^3 \, p^0}\right] 
\left[{d^3 k \over (2\pi)^3 \, k^0}\right] 
\sqrt {(p\cdot k)^2 - m_p^2 \, m_k^2} \, \rho_p \, \rho_k \, d\sigma. 
\label{etadsig}
\end{equation}

To get the rate of collision per unit time
$\eta^{(1)}$ of one $p$-particle in the 
plasma, we divide $\eta$ by $\rho_p \, d^3 p/(2\pi)^3$, the 
volume density of $p$-particles:
\begin{equation}
\eta^{(1)} = {d^3 k \over (2\pi)^3 \, p^0 k^0 } 
\sqrt {(p\cdot k)^2 - m_p^2 \, m_k^2} \, \rho_k \, d\sigma.
\label{eta1}
\end{equation}
To calculate the mean free path associated with the above process,
 we need to find the total 
rate of collision per unit time $\eta_{tot}^{(1)}(p)$ for one
particle with initial 4-momentum $p$ into any final state in the
allowed phase space, using the
total cross section $\sigma$.  From Eqs.\ (\ref{dsig}) and (\ref{eta1}) we get
\begin{equation}
\eta_{tot}^{(1)}(p) = {2 \over (4 \pi)^5} \int 
{d^3 k \over p^0 k^0 } 
\int {d^3 p^\prime d^3 k^\prime \over p^{\prime 0} k^{\prime 0}} 
\delta^{(4)}(p + k - p^\prime - k^\prime)|{\cal M}|^2 \, \rho_k
(1 \pm \rho_{p^{\prime}}) (1 \pm \rho_{k^{\prime}}).
\label{eta1totp}
\end{equation}
Note that each scatterer included will give an additive contribution of
this form to $\eta_{tot}^{(1)}$.

The collision time $\tau(p)$, the length of time between two successive
collisions for a $p$-particle, is the inverse of $\eta_{tot}^{(1)}(p)$
\begin{equation}
\tau(p) = {1 \over \eta_{tot}^{(1)}(p)}.
\label{taup}
\end{equation}
The distance $l(p)$ such a $p$-particle travels between two
collisions is then given by $l(p) = (|\vec p|/p^0) \, \tau(p)$.  We finally get
the mean free path $\lambda$ for the $p$-particle by taking the thermal average
of $l(p)$, using the thermal distribution of the $p$-particles.  We thus get
\begin{equation} 
\lambda = \left[\int \frac{d^3 p}{(2 \pi)^3} \, \rho_p
\right]^{- 1} \int \frac{d^3 p}{(2 \pi)^3} \, \rho_p {|\vec p| \over p^0 \,
\eta_{tot}^{(1)}(p)} \, ,
\label{lambda}
\end{equation}
where we have used Eq.\ (\ref{taup}).  Note that this mean free path vanishes
if the cross section suffers from infrared divergences.  However, for diffusive
processes, these divergences are suitably regulated and only processes with 
nontrivial momentum transfer contribute.  The resulting 
mean free path (\ref{lambda}) can then be related to the diffusion constant $D$
by the relation
\begin{equation} 
D = \frac{1}{3} \, \lambda \, {\bar v}, 
\label{D}
\end{equation} 
where ${\bar v}$ is the mean velocity of the diffusing particle.

\section{Results and Discussion}

We begin this section by describing some of the thermal properties of gluons
in the plasma and how we incorporate these properties into our calculations. 
Due to interactions with the plasma, gluons develop temperature dependent
masses.  The longitudinal gluons have a thermal Debye mass $m_g(T) = \sqrt {8
\pi \alpha_s} \, T$, where $\alpha_s = g_s^2/(4 \, \pi)$, at the 1-loop level
and the transverse gluons have a non-perturbative thermal magnetic mass
$m_t(T)$ that is zero at the 1-loop level and is expected to be ${\cal
O}(g_s^2 \, T)$.  Thus, we may assume that the infrared screening of
longitudinal gluons occurs at a momentum scale $m_g$, and the similar scale
for the transverse gluons is likely lower.  At the electroweak phase
transition temperature $T_c \approx 100$ GeV, $\alpha_s \approx 0.1$ and $m_g
\approx 1.6 \, T$.  Since the magnetic mass is unknown, the choice of
transverse infrared momentum cutoff is rather arbitrary.  However, because
$g_s \approx 1$ at scale $T_c$, it is reasonable to assume that $m_t$ is of
order $T$.  In Table \ref{tbl1}, we take as two representative values $m_t =
T$ and $m_t = m_g$.

\begin{figure}\begin{center}
\begin{picture}(25000,21000)
\drawline\scalar[\SW\REG](12500,7000)[4]
\put(7500,3500){$p$}
\put(8500,4500){\vector(1,1){1500}} 
\drawline\scalar[\SE\REG](12500,7000)[4]
\put(17500,3500){$p'$}
\put(15000,6000){\vector(1,-1){1500}} 
\drawline\gluon[\N\REG](12500,7000)[6]
\drawline\fermion[\NW\REG](\gluonbackx,\gluonbacky)[8000]
\put(7500,16500){$k$}
\put(8500,16000){\vector(1,-1){1500}} 
\drawline\fermion[\NE\REG](\gluonbackx,\gluonbacky)[8000]
\put(17000,16500){$k'$}
\put(15000,14500){\vector(1,1){1500}} 
\end{picture}
\end{center}
\caption{$t$-channel squark-quark Feynman diagram.}
\label{t-sq}
\end{figure}
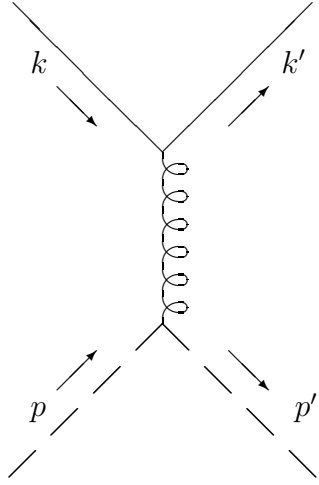

The amplitude for the $t$-channel squark-quark diagram of Fig.\ \ref{t-sq} is
\begin{equation}
{\cal M}_{sq} = - 8 \pi \, \alpha_s \, T^a \,  p^\mu D_{\mu \nu} \, 
{\bar q} (k^\prime) \, T^a \gamma^\nu q (k),
\label{Msq}
\end{equation}
where $T^a$ is a generator in the adjoint
representation of the $SU(3)_c$ color gauge group, and $q$ is a quark
spinor.  We work in the Landau gauge where, as explained in the
appendix, $(k^\prime - k)^\mu D_{\mu \nu} = 0$.  Since $p + p^\prime =
2p + k - k^\prime$, ${\cal M}_{sq}$ does not depend on $p^\prime$
explicitly.  We use the amplitude ${\cal M}_{sq}$ of Eq.\ (\ref{Msq})
to estimate the squark mean free path $\lambda_s$ and diffusion
constant $D_s$ from Eqs.\ (\ref{lambda}) and (\ref{D}).  The amplitude
is squared and summed over all $72$ species of quark scatterers ($3$
colors, $2$ spins, $6$ flavors, and antiparticles).  We have numerically 
computed $\lambda_s$ and $D_s$ for supersymmetry breaking squark masses 
ranging from 50 GeV to 200 GeV and observed that their mass dependence is 
weak.  Our results are presented in Table
\ref{tbl1}, where the values of $\lambda_s$ and $D_s$ have been computed 
for $m_s = 100$ GeV.

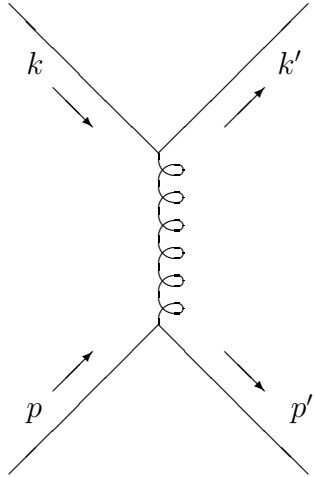
\begin{figure}
\begin{center}
\begin{picture}(25000,21000)
\drawline\fermion[\SW\REG](12500,7000)[8000]
\put(7500,3500){$p$}
\put(8500,4500){\vector(1,1){1500}} 
\drawline\fermion[\SE\REG](12500,7000)[8000]
\put(17500,3500){$p'$}
\put(15000,6000){\vector(1,-1){1500}} 
\drawline\gluon[\N\REG](12500,7000)[6]
\drawline\fermion[\NW\REG](\gluonbackx,\gluonbacky)[8000]
\put(7500,16500){$k$}
\put(8500,16000){\vector(1,-1){1500}} 
\drawline\fermion[\NE\REG](\gluonbackx,\gluonbacky)[8000]
\put(17000,16500){$k'$}
\put(15000,14500){\vector(1,1){1500}} 
\end{picture}
\end{center}
\caption{$t$-channel quark-quark Feynman diagram.}
\label{t-qq}
\end{figure}
In Ref.\ \cite{JPT}, $D_q$ is computed using only the $t$-channel
quark-quark scattering amplitude of Fig.\ \ref{t-qq}, given by
\begin{equation}
{\cal M}_{qq} = - 4 \pi \, \alpha_s \, {\bar q} (p') \, T^a \gamma^\mu q(p) \,
D_{\mu \nu} \, {\bar q} (k^\prime) \, T^a \gamma^\nu q(k).
\label{Mqq}
\end{equation}

To compare our method with that of Ref.\ \cite{JPT}, we have computed $D_q$,
again numerically, using only ${\cal M}_{qq}$ (and again summing over $72$
species of scatterers).  The entries in Table \ref{tbl1} labeled ``JPT'' refer
to the numbers we get for the set of parameters that are used in Ref.\
\cite{JPT}, namely $\alpha_s = 1/7$ and $m_g = m_t = \sqrt{8 \pi \alpha_s} \,
T = 1.9 \, T$.  We see that using the JPT parameters, we obtain the estimate
$D_q \sim 6/T$ of Ref. \cite{JPT}, where only the leading logarithmic
contributions were considered.  This suggests that our momentum randomization
approximation, where the momentum transfer $q \grtsim T$, yields reasonable
estimates for the diffusion constant.  However, we also note that in Ref. 
\cite{JPT} the diffusion constant $D \sim 1/[g_s^4 \ln (g_s^{-1})]$, where the logarithmic dependence comes from the use of a $g_s$ dependent $m_g$
as the infrared regulator.  In our approach, as mentioned before, for physical
values of $g_s$, we can use $m_g$, and perhaps $m_t$, as the infrared cutoffs,
but in general, the cutoff we choose is not a function of $g_s$, and does not
vanish as $g_s \to 0$, for it has to be chosen to satisfy $q \grtsim T$.  In
this way, our computations will yield $D \sim 1/(g_s^4)$, which represents the
$\alpha_s$ dependence of the amplitudes (\ref{Msq}) and (\ref{Mqq}). 
Therefore, the behavior of our results differs from those of Ref. \cite{JPT}
by $1/\ln(g_s^{-1})$.  However, for physical values of $g_s$, the logarithm is
of order unity.  Taking the result of Ref. \cite{JPT} as a fair
estimate of $D_{q}$, we believe that our results are reliable up to factors
of order unity.  Note that we do not consider all the processes that
contribute at this level: scatterings from on-shell gluons in the plasma also
provide a substantial contribution.  However, the results of Ref.\
\cite{Moore} suggest that the inclusion of these diagrams will not change the
results by more than a factor of 2.

\begin{table} 
\begin{tabular}{|c||c|c|c|c|c|l|}
& $\alpha_s$ & $m_g$ & $m_t$ & $\lambda$ & $D$ & Remarks \\ \hline \hline 
quark & $1/10$ & $1.6 \, T$ & $T$ & $13/T$ & $4/T$ &  \\ 
& $1/10$ & $1.6 \, T$ & $1.6 \, T$ & $24/T$ & $8/T$ &  \\ 
& $1/7$ & $1.9 \, T$ & $1.9 \, T$ & $18/T$ & $6/T$ & JPT \\ \hline 
squark & $1/10$ & $1.6 \, T$ & $T$ & $12/T$ & $3/T$ &  \\ 
$m_s = 100$ GeV & $1/10$ & $1.6 \, T$ & $1.6 \, T$ & $18/T$ & $5/T$ & \\ 
& $1/7$ & $1.9 \, T$ & $1.9 \, T$ & $14/T$ & $4/T$ & JPT \\ 
\end{tabular} 
\caption{Results for $\lambda$ and $D$ } 
\label{tbl1} 
\end{table}

Our calculations suggest that $\lambda_s \approx \lambda_q$ and $D_s \approx
D_q$, up to factors of order unity, and most likely to within $30\%$, and that
the effects of different statistics, masses, and couplings on the values of
$D$ and $\lambda$ for squarks and quarks are not strong.  For $m_t = T < m_g$,
we roughly get $\lambda \lsim 10/T$ and $D \lsim 3/T$.  On the other hand, if
$m_t = m_g$, our results increase by about a factor of 2.  

Electroweak baryogenesis scenarios that use quarks or squarks for charge
transport in a first order phase transition suffer from a suppression due to
decoherence that is caused by multiple scatterings of these strongly
interacting particles across the width of the expanding Higgs wall.  A measure
of the strength of decoherence is the ratio of the mean free path of the
particles to the width $w$ of the wall\cite{DRW}.  In Ref.\ \cite{Quiros}, a
2-loop MSSM calculation of the Higgs wall profile gives $w \approx 25/T$. 
Using this value of $w$, our results yield $\lambda/w \lsim 1/2$, which
suggests that although the effects of decoherence are not negligible, they are
not severe.  
       
\section*{Acknowledgements}

It is a pleasure to thank Ann Nelson, John Preskill, Krishna Rajagopal, Dam
Son, Mark Wise, and Laurence Yaffe for insightful discussions.  We would also
like to thank Martin Gremm and Iain Stewart for their helpful comments.  This
work was supported in part by the U.S.\ Dept.\ of Energy under Grant No.\
DE-FG03-92-ER40701.

\appendix
\section*{Thermal Gluon Propagator}

In this appendix, we give the expression we use for the approximate
thermal gluon propagator in a plasma, taking the different properties
of the longitudinal and transverse gluons into account as represented
by their respective cutoffs $m_g$ and $m_t$.  Let $n^\mu = (1, 0, 0,
0)$ be the 4-velocity of the plasma in the plasma frame.  We denote
the 4-momentum $q$ of the propagating gluon by $q^\mu=(q^0,\vec{q})$
in the plasma frame.  The component of $n$ that is orthogonal to $q$
is given by ${\tilde n}$, where
\begin{equation}
{\tilde n}^\mu = n^\mu - \frac{(n \cdot q) q^\mu}{q^2}.
\label{ntil}
\end{equation}
We define two projection operators $P_{\mu \nu}$ and $Q_{\mu \nu}$, where
\begin{equation}
P_{\mu \nu} = g_{\mu \nu} - \frac{q_\mu q_\nu}{q^2} - \frac{{\tilde n}_\mu
{\tilde n}_\nu}{{\tilde n}^2}
\label{P}
\end{equation} 
and
\begin{equation}
Q_{\mu \nu} =\frac{{\tilde n}_\mu {\tilde n}_\nu}{{\tilde n}^2}.             
\label{Q}
\end{equation}

The expression for the Landau gauge thermal gluon propagator in our 
approximation is then given by
\begin{equation}
D^{(L)}_{\mu \nu} = - \left[\frac{1}{q_T^2} P_{\mu \nu} + 
\frac{1}{q_L^2} Q_{\mu \nu}\right],
\label{Dapp}
\end{equation}
where $q_T^2 = q^2 - m_t^2$ and $q_L^2 = q^2 - m_g^2$, 
and from $q^\mu P_{\mu \nu} = q^\mu Q_{\mu \nu} = 0$, 
we have $q^\mu D^{(L)}_{\mu \nu} = 0$.

\end{document}